\documentclass[11pt]{article}
\pdfoutput=1
\usepackage{setspace,braket}
\usepackage{amsmath, amssymb}
\usepackage{etoolbox}
\usepackage{jheppub}
\usepackage{array,tabularx}
\usepackage{hyperref}

    \makeatletter
    \patchcmd{\maketitle}{\@fpheader}{}{}{}
\makeatother

\newcommand{\no}[1]{:\mathrel{\mkern2mu #1 \mkern2mu}:}
\newcommand{\ZZ}{\mathbb{Z}} 

\def\W{{\mathcal W}}

\def\be{\begin{equation}}
\def\ee{\end{equation}}

\def\a{\alpha}
\def\b{\beta}
\def\g{\gamma}
\def\d{\delta}

\def\l{\lambda}
\def\m{\mu}\def\n{\nu}\def\l{\lambda}

\def\bg{\bar{g}}

\def\beq{\begin{eqnarray}}\def\eeq{\end{eqnarray}}
\def\ba#1\ea{\begin{align}#1\end{align}}
\def\bg#1\eg{\begin{gather}#1\end{gather}}
\def\bm#1\em{\begin{multline}#1\end{multline}}
\def\bmd#1\emd{\begin{multlined}#1\end{multlined}}

\def\a{\alpha}
\def\b{\beta}

\def\d{\mathfrak{d}}

\def\g{\gamma}

\def\l{\lambda}
\def\L{\Lambda}
\def\m{\mu}
\def\n{\nu}
\def\p{\phi}

\def\nn{\nonumber}

\def\({\left(}
\def\){\right)}
\def\[{\left[}
\def\]{\right]}

\def\Tr{{\rm Tr}}

\def\N{{\mathcal N}}

\def\half{{\frac{1}{2}}}

\newcommand{\equ}[1]{Eq.~(\ref{#1})}
\newcommand{\sct}[1]{Sec.~\ref{#1}}

\renewcommand\arraystretch{1.5}

\begin{document}
\title{Symmetry Algebras of Stringy Cosets}
\author[a]{Dushyant Kumar}
\author[b]{Menika Sharma}
\affiliation[a]{Harish-Chandra Research Institute, Jhusi, Allahabad, India}
\emailAdd{sehrawat.dushyant@gmail.com}
\affiliation[b]{Department of Mathematics, City, University of London,
Northampton Square, EC1V 0HB London, UK}
\emailAdd{menika.sharma@city.ac.uk}
\begin{abstract}{We find the symmetry algebras of cosets which are generalizations of the minimal-model cosets, of the specific form $\frac{SU(N)_{k} \times SU(N)_{\ell}}{SU(N)_{k+\ell}}$. We study this coset in its free field limit, with $k,\ell \rightarrow \infty$, where it reduces to a theory of free bosons. We show that, in this limit and at large $N$, the algebra $\W^e_\infty[1]$ emerges as a sub-algebra of the coset  algebra.  The full coset algebra is a larger algebra than conventional $\W$-algebras, with the number of generators rising exponentially with the spin, characteristic of a stringy growth of states. We compare the coset algebra to the symmetry algebra of the large $N$ symmetric product orbifold CFT, which is known to have a stringy symmetry algebra labelled the `higher spin square'. We propose that the higher spin square is a sub-algebra of the symmetry algebra of our stringy coset.}
\end{abstract}
\toccontinuoustrue
\maketitle
\allowdisplaybreaks[1]

\section{Introduction}

The papers \cite{Gaberdiel:2012} established a duality between the CFT of the coset model
\be
\label{specialcoset}
	\frac{SU(N)_{1} \times SU(N)_{\ell} }{SU(N)_{1+\ell}}\,,
\ee
and three-dimensional higher-spin Vasiliev theory \cite{vas1}, in the large $N,\ell$ limit. This duality is characterized by a large symmetry algebra  $W_\infty[\mu]$  on the CFT side which is interpreted as the asymptotic symmetry algebra on the bulk side \cite{Campoleoni:2010}. The parameter $\mu=\frac{N}{\ell+N}$ is the 't~Hooft coupling on the CFT side, while on the bulk side it determines the mass of the scalar field.  The $\W_\infty[\mu]$ algebra consists of generators with spin $2$ to $\infty$ with each generator having multiplicity one and the commutation relations of these generators depend on the parameter $\mu$.

Our aim is to find the symmetry algebra for the coset
\be
\label{gencoset}
	\frac{SU(N)_{k} \times SU(N)_{\ell} }{SU(N)_{k+\ell}}\,
\ee
which is a generalization of the coset in \equ{specialcoset}.
The central charge for this coset is
\be
	\frac{k \ell (k + \ell + 2 N) (N^2 - 1)}{(k + N) (\ell + N) (k + \ell + N)}\,.
\ee
The coset in \equ{gencoset} has three independent parameters and two interesting limits. The first limit arises on taking $N$ and $\ell$ to infinity while holding $k$ and $\mu=\frac{N}{N+\ell}$ fixed. The central charge reduces to
\be
	\frac{k ( N^2-1)}{k + N} \sim k N (1- \mu^2)\,.
\ee
Since, the central charge scales as $\sim N$, in this limit the coset in \equ{gencoset} is usually referred to as a vector coset model. This vector coset model was studied in Refs.~\cite{Creutzig:2013}, where a related, but different, coset model $SU(k+\ell)_N/SU(\ell)_N$  was proposed as the CFT dual to a bulk Vasiliev theory with a matrix extension.

There exist other limiting procedures which result in a central charge of the coset in \equ{gencoset} which scales as $N^2$. One  way is to take $N$ and $k, \ell$ to infinity while holding $k-\ell$ and $\mu=\frac{N}{k+\ell+N}$ fixed. In this case, the central charge scales as $\sim N^2(1- \frac{1}{\mu})$.
A variation of this, which is the limiting procedure we use in this paper, is to take $N$ and $k, \ell$ to infinity with $N/\ell$ set to zero and 
\be 
\label{lambdac}
	\lambda=\frac{N}{k}
\ee
fixed. The coset central charge is then
\be
	\frac{N^2-1}{1+\lambda}\,.
\ee
Since the central charge scales as a matrix model, in this limit we expect the coset in \equ{gencoset} to have a string dual and we refer to it in the text as the stringy $SU(N)$ coset. In this paper, we will study the symmetry algebra of this stringy $SU(N)$  coset in the limit where $k$ and $\ell$ go to infinity, but keep $N$ finite.  Thus, we are not determining the algebra explicitly at large $N$. However, from the general behavior of the algebra at finite $N$, we can infer many of its properties at infinite $N$ which we will elaborate on in the text. In particular, this method tells us the properties of the large $N$ coset algebra at $\lambda=0$ with $\lambda$ defined as in \equ{lambdac}. 

Historically, the algebra for the coset model in \equ{specialcoset} was first studied at finite $N$ before the infinite $N$ case was dealt with. The finite $N$ algebra is called $\W_N$ \cite{Zamolodchikov:1985} and has generators ranging from $2$ to $N$ with multiplicity one. Indeed, it has taken many years to completely understand the large $N$ limit of the $\W_N$ algebra \cite{Bakas:1990, Gaberdiel:2012t, Linshaw:2017}. 
The algebra of the coset in \equ{gencoset} for small $N$ and large $k,\ell$ has been studied before in Refs. \cite{Bouwknegt:1992wg,Bais:1987a,Bais:1987b}, although it has only attracted a fraction of the attention that the $\W_N$ algebra has and perhaps rightly so. $\W_N$ algebras, which are extensions of the Virasoro algebra, have complicated commutation relations but a simple spectrum of fields. In contrast, the symmetry algebras of the coset theories in \equ{gencoset} have a spectrum of generators with the multiplicity climbing at a exponential rate with the spin (the algebra still has, of course, a finite number of generators  at finite $N$). Unlike their $\W_N$ counterparts, these algebras belong to a class of algebras which are finitely non-freely generated \cite{deBoer:1993} and thus are less tractable. On the other hand, the fact that this coset model and supersymmetric generalizations have generators whose multiplicity increases with spin makes them prime candidates to be dual to string theories in AdS. It is with this motivation that we study them in this paper. 

We exclusively work with the bosonic coset in \equ{gencoset}, so that we can study the symmetry algebra in its simplest form. A $\N=2$ supersymmetric generalization of the coset in \equ{gencoset} was studied in \cite{Gopakumar:2012}. Related work for a coset with $\N=1$ supersymmetry appears in Refs.~\cite{Ahn:1990,Ahn:2012}. However, a crucial distinction between our analysis and the supersymmetric cases studied is that we are working in the limit of zero coupling, with a free theory.

The coset theories that we study are similar to $SU(N)$ gauge theories in four dimensions which are known to have string duals on the $AdS_5$ background. However, string theories on $AdS_3$ are expected to be dual to a different family of CFTs: symmetric product orbifolds. In this paper, we explore the relation between the symmetry algebra of the bosonic symmetric product orbifold theory and the coset theory. To be able to do this, we explicitly write down the currents of the coset theory. For the coset in \equ{specialcoset} with level $k=1$, the currents of the $\W$-algebra correspond to Casimir operators of $SU(N)$. For the more general coset theory, currents of the $\W$-algebra can be generated from the Casimir operators by sprinkling additional derivatives on the constituent currents. We construct these currents in Section \ref{secCurrents}. However, as we will see, the coset theory also has additional currents that cannot be constructed from the Casimir operators. 

Information about the generators of any coset theory resides in the vacuum character of the partition function of the theory. In Section \ref{secGrowth}, we write down the vacuum character of the coset in \equ{gencoset}. Unlike the case $k=1$, it is not possible to formulate this character in closed form for general $k$ and $N$ and it can only be expressed in terms of string functions. We therefore resort to numerical techniques to find the generators of the algebra from its vacuum character for low values of $N$, following Ref.~\cite{Bouwknegt:1992wg}. Later in Section \ref{secCurrents}, when we explicitly construct the currents for finite $N$, the calculation in Section \ref{secGrowth} serves as a touchstone for our results. 

This paper is organized as follows. In \sct{secGrowth} we find the low lying spectrum of the symmetry algebra of the coset in \equ{gencoset} for small values of $N$ in the large $k,\ell$ limit. In \sct{secCurrents} we construct the currents for this same coset for the special values $N=2$ and $N=3$. The $N=3$ case is especially important for understanding the structure of the currents at general $N$ and we present this case in some detail. In \sct{secCurrents}, we also work out the relation of the coset algebra to the algebra $\W_\infty[\mu]$ and also its relation to the higher spin square. The algebra of the symmetric product orbifold at general values of $N$ is worked out in Appendix~C.

\section{Perturbative growth of states} \label{secGrowth}
In this section, we compute the vacuum character of the coset model in \equ{gencoset} at finite $N$, with $k,\ell \rightarrow \infty$. This computation will tell us at what rate the perturbative states of the current algebra grow with the spin.
The density of states of a  CFT partition function in the regime of large spin $s$ but $s< c$ determines the dual holographic theory.  Since, in this paper, we are only interested in the symmetry algebra of the coset in \equ{gencoset}, we will focus on the vacuum character. We will not be able to determine the vacuum character exactly but will compute the low-lying spectrum of the symmetry algebra. We carry this out in Sec.~\ref{growthN} and our results appear in Table~\ref{t:1}. It is also of interest to compute the vacuum character at fixed $k$ at finite $N$, since this helps us to understand the nature of the symmetry algebra of our stringy coset. We do this in Sec.~\ref{growthk} and the results appear in Tables~\ref{t:2} and ~\ref{t:3}. In Sec.~\ref{asymptoticN}, we determine the asymptotic growth of states of the vacuum character.  In the following, we describe the method we use to compute the vacuum character. 

It is well known that the $\W$-algebra of the coset $\mathfrak{g}_k/\mathfrak{g}$ is the same as the $\W$-algebra of the coset 
 $(\mathfrak{g}_k \oplus \mathfrak{g}_\ell)/\mathfrak{g}_{k+\ell}$ in the $\ell \rightarrow \infty$ limit. Therefore, to find the symmetry algebra for the coset 
\be
	\frac{SU(N)_{k} \otimes SU(N)_{\ell}}{SU(N)_{k+\ell}}
\ee
as $\ell \rightarrow \infty$, we find the algebra of the coset model 
\be
\label{coset2}     
	\frac{SU(N)_{k}}{SU(N)}\,.
\ee
We will use the so-called ``character technique''. To find the coset symmetry algebra we need to look at the vacuum character, so we work out the branching function $b_\l^\L(q) $ for the weights: $\L=(k,0,\cdots)$ and $\l=(0,0,\cdots)$. This will give us a series in the variable $q$. We can rearrange this series as 
\be
\label{fvac}
	\frac{q^{-\frac{c}{24}}}{\prod_{i=1}^{l} F_{s_i}(q)}(1- j q^n \cdots)\,,
\ee
where
\be
 	F_{s} \equiv \prod_{k=s}^{\infty} (1-q^k)\,.
\ee
\equ{fvac} is the general form of the vacuum character for an algebra with fields of spins $s_i$, where $i$ ranges from $1$ to $l$ and with $j$ null states starting at order $n$. Here, $l$ is the total number of generators of the algebra. Note that the character-technique is not fool-proof. The actual algebra may have additional currents, since we can always add currents to the denominator of \equ{fvac}, while at the same time increasing the number of null states to keep the vacuum character unchanged. Nevertheless,  studying the vacuum character gives us a good indication of the nature of the algebra. We also restrict our attention to the vacuum character and ignore any extensions of the coset algebra at specific values of the level $k$.

The branching function for the coset in \equ{coset2} is given by
\be
\label{branching1}
	b_\l^\L(q) \equiv {\rm Tr\,}_{L_{\L,\l}}\ q^{L_0 -c/24}
	= \sum_{w\in W} \epsilon(w) c^\L_{w(\l+\rho)-\rho + k\L_0}(q)
	q^{ {\frac{1}{2k}} |w(\l+\rho)-\rho|^2}\,,
\ee
where the $c^\L_\l(q)$ are the Kac-Peterson string functions defined in \equ{stringfunction}. 

\subsection{ Algebra for small $N$ }\label{growthN}
We now take $k$ to $\infty$. Then the branching function in \equ{branching1} reduces to
\be
\label{branching2}
	b_\l^\L(q) = \sum_{w\in W} \epsilon(w) c^\L_{w(\l+\rho)-\rho + k\L_0}(q)\,.
\ee
The string functions are given by
\be
\label{stringfunction}
c^\L_\l(q) = {q^{-c/24}\over \prod(1-q^n)^{{\rm dim\,}\mathbf{g}} }
\sum_{w\in \widehat{W}} \epsilon(w) q^{h_{w*\L,\l} }
\sum_{ \{ n_\b\in\ZZ | \sum_{\b\in\Delta_+} n_\b\b = w*\L -\l\} }
\left(\prod_{\b\in\Delta_+} \p_{n_\b}(q) \right)\,,
\ee
where we have introduced
\be
\p_n = \sum_{m\geq0} (-1)^m q^{\half m(m+1) +nm}\,,\qquad\qquad
\p_{-n}(q) = q^n \p_n(q)\,,
\ee
\be
h_{\L,\l} = { (\L,\L+2\rho) \over 2(k+h^V)} - {(\l,\l)\over 2k}\,,
\ee
and $w * \L = w(\L+ \rho) - \rho$. Here $\Delta_+$ denotes the set of positive roots of $\mathfrak{g}$.
In the large $k$ limit, the sum in \equ{stringfunction} over the affine Weyl group elements will reduce to a sum over the finite Weyl group elements. Thus the expression in \equ{stringfunction} simplifies to
\be
\label{stringfunction1}
	c^\L_\l(q) = {q^{-c/24}\over \prod(1-q^n)^{{\rm dim\,}\mathbf{g}} }
\sum_{w\in W} \epsilon(w) 
\sum_{ \{ n_\b\in\ZZ | \sum_{\b\in\Delta_+} n_\b\b = w*\L -\l\} }
\left(\prod_{\b\in\Delta_+} \p_{n_\b}(q) \right)\,.
\ee

\subsubsection{Algebra for stringy $SU(2)$ coset } 
In this case, there are two Weyl group elements: $1$ and $w_{\alpha_1}$. Thus, the branching function in \equ{branching2} becomes
\be
	b_{(0)}^{(k,0)}(q) =c^{(k,0)}_{(k,0)}(q) - c^{(k,0)}_{(k-2,2)}(q)  \,.
\ee
In the limit $k\rightarrow \infty$, the string functions appearing in the RHS of the above expression are given by
\be
	c^{(k,0)}_{(k,0)}(q) = {q^{-c/24} \over \prod(1-q^n)^{3} } \big\{\phi_0(q) - \phi_{-1}(q) \big \} 
\ee
and
\be
	c^{(k,0)}_{(k-2,2)}(q) = {q^{-c/24}\over \prod(1-q^n)^{3} } \big\{\phi_{-1}(q) - \phi_{-2}(q) \big\}\,.
\ee
These expressions can be read off from \equ{stringfunction1}. Working out (and rearranging) the branching function we get
\be
	b_{(0)}^{(k,0)}(q) = \frac{1}{F_2 F_4 F_6^2 F_8^2 F_9 F_{10}^2 F_{12}}(1-q^{13} -3 q^{14} -7 q^{15}- \cdots)
\ee
Thus the coset $\frac{SU(2)_k}{SU(2)}$ has a symmetry algebra with generators of spin 
\be
\label{SU2currents}
	2,4,6^2,8^2,9,10^2, 12
\ee
in the large $k$ limit. 

\subsubsection{Algebra for stringy $SU(3)$ coset}
The branching function in \equ{branching2} for the case of the coset $\frac{SU(3)_k}{SU(3)}$ becomes
\be
	b_{(0,0)}^{(k,0,0)}(q) =c^{(k,0,0)}_{(k,0,0)}(q) - 2 c^{(k,0,0)}_{(k-2,1,1)}(q) + c^{(k,0,0)}_{(k-3,3,0)}(q)+ c^{(k,0,0)}_{(k-3,0,3)}(q)   - c^{(k,0,0)}_{(k-4,2,2)}(q)  \,.
\ee
Defining 
\be
	\zeta(q)= q^{-c/24} \frac{1}{\prod(1-q^n)^{8} }\,,
\ee
and calculating the string functions in the limit $k\rightarrow \infty$ we find
\begin{align}
c^{(k,0,0)}_{(k,0,0)}(q) &=\zeta(q) \sum_{n\in\mathbb{Z}}\phi_{-n}(q)  \big\{ 2\phi_{n-1}(q)\phi_{n-2}(q)-2\phi_{n-1}(q)\phi_{n}(q)-\phi_{n-2}(q)^2+\phi_{n}(q)^2  \big\}\,,\nonumber \\
c^{(k,0,0)}_{(k-2,1,1)}(q) &=\zeta(q) \sum_{n\in\mathbb{Z}}\phi_{-n}(q)  \big\{ 2\phi_{n-2}(q)\phi_{n-3}(q)-2\phi_{n-1}(q)\phi_{n-2}(q)-\phi_{n-3}(q)^2+\phi_{n-1}(q)^2  \big\}\,,\nonumber\\
c^{(k,0,0)}_{(k-3,3,0)}(q) & = \zeta(q)\sum_{n\in\mathbb{Z}}\phi_{-n}(q)  \big\{\phi_{n-4}(q)\phi_{n-2}(q)-\phi_{n-4}(q)\phi_{n-3}(q)-\phi_{n-3}(q)\phi_{n-1}(q) \nonumber\\[-3\jot]  &~~~~~~~~~~~~~~~~~~~~~~~~~~~~~~~~~~~~~~~~~~~~~+\phi_{n-3}(q)^2+ \phi_{n-2}(q)\phi_{n-1}(q)-\phi_{n-2}(q)^2  \big\}\,,\nonumber\\
\nonumber\\
c^{(k,0,0)}_{(k-4,2,2)}(q) &= \zeta(q) \sum_{n\in\mathbb{Z}}\phi_{-n}(q)  \big\{2\phi_{n-3}(q)\phi_{n-4}(q)-2\phi_{n-2}(q)\phi_{n-3}(q)-\phi_{n-4}(q)^2+\phi_{n-2}(q)^2  \big\}\,.\nonumber\\
\end{align}
In addition,
\be
	c^{(k,0,0)}_{(k-3,0,3)}(q)  = c^{(k,0,0)}_{(k-3,3,0)}(q)\,.
\ee
\begin{table}[!b]
 \begin{center}
\renewcommand{\arraystretch}{1.5}
 \begin{tabular}{  c l l}
 \hline
 \hline
 $N$& Vacuum Character& Algebra\\
 \hline
 \hline
 $2$ &$1 + q^2 + q^3 + 3 q^4 + 3 q^5 + 8 q^6 + 9 q^7 + 19 q^8 + \cdots$&$2, 4, 6^2,8^2,\cdots\,.$ \\
 \hline
 $3$ &$1 + q^2 + 2 q^3 + 4 q^4 + 6 q^5 + 15 q^6 + 22 q^7 + 46 q^8 + \cdots$&$2, 3, 4, 5, 6^4, 7^2, 8^7,\cdots$ \\
 \hline
  $4$ &$1 + q^2 + 2 q^3 + 5 q^4 + 7 q^5 + 18 q^6 + 29 q^7 + 64 q^8 + \cdots$ &$2, 3, 4^2, 5,  6^5, 7^4 , 8^{12},\cdots $\\
  \hline
  $5$ &$1 + q^2 + 2 q^3 + 5 q^4 + 8 q^5 + 19 q^6 + 32 q^7 + 71 q^8 + \cdots$ &$2, 3, 4^2, 5^2,  6^5, 7^5, 8^{14},\cdots $\\
  \hline
  $6$ &$1 + q^2 + 2 q^3 + 5 q^4 + 8 q^5 + 20 q^6 + 33 q^7 + 74 q^8+ \cdots$ & $2, 3, 4^2, 5^2, 6^6, 7^5, 8^{15},\cdots $ \\
  \hline
  $7$&$1 + q^2 + 2 q^3 + 5 q^4 + 8 q^5 + 20 q^6 + 34 q^7 + 75 q^8+ \cdots$ & $2, 3, 4^2, 5^2, 6^6, 7^6, 8^{15},\cdots $\\
  \hline
  \hline
 \end{tabular}
 \caption{ \label{t:1}The vacuum character for the stringy coset model for small values of $N$. The corresponding algebra appears in the third column. The central charge of the coset is related to $N$ by $c=N^2-1$. Note that the vacuum character (and hence the algebra) stabilizes till order $q^N$: which means that the generators up to spin $N$ do not change on further increasing $N$.}
  \end{center}
 \end{table}
Working out the branching function we get
\be
\label{n3char}
	b_{(0)}^{(k,0)}(q) = \frac{1}{F_2 F_3 F_4 F_5 F_6^4  F_7^2 F_8^7  F_9^9 F_{10}^{12}  F_{11}^{16}  F_{12}^{26}  F_{13}^{26}  F_{14}^{33}  F_{15}^{33}  F_{16}^{12}  }(1 - 24 q^{17} -137 q^{18}- 404 q^{19}-\cdots)\,.
\ee
The spin of the generators of the algebra and their multiplicity can now be read off from the denominator. Our results for the $N=2$ and $N=3$ cases agree with those in Ref.~\cite{Bouwknegt:1992wg}.

We can work out the vacuum character for the coset $\frac{SU(4)_k}{SU(4)}$, in a similar fashion, and we find that the algebra has generators of spin:
\be
\label{VCSU4}
 2, 3, 4^2, 5, 6^5, 7^4, 8^{12}, 9^{15},  10^{28}, 11^{41}, 12^{75}, 13^{103},  14^{166},  15^{235},  16^{313},  17^{362},  18^{310}. 
\ee
Using Mathematica, we have worked out the symmetry algebra for the coset $\frac{SU(N)_k}{SU(N)}$ in the large $k$ limit till $N=7$ and up to generators of spin $8$. The results appear in Table~\ref{t:1}.  As we can see from the table, the currents up to spin $N$ stop changing as $N$ is further increased. We, therefore, expect the algebra at $N=\infty$ to have the following low-lying spectrum of generators:
\be
\label{VCSUN}
	2,3,4^2,5^2,6^6,7^6,\cdots\,.
\ee

\subsection{Algebra for cosets with finite $k$  }\label{growthk}
\begin{table}[!b]
\begin{center}
\renewcommand{\arraystretch}{1.5}
\resizebox{0.9 \textwidth}{!}{
 \begin{tabularx}{\textwidth}{c X l}
 \hline
 \hline
 $N$& Vacuum Character& Algebra\\
 \hline
 \hline
 $2$ &$1 + q^2 + q^3 + 3 q^4 + 3 q^5 + 8 q^6 + 9 q^7 + 19 q^8 + 25 q^9 + 
 45 q^{10} + 61 q^{11} + 105 q^{12}+ \cdots$ & $2,4, 6^2, 8,9\,.$\\
 \hline
 $3$ &$1 + q^2 + 2 q^3 + 4 q^4 + 6 q^5 + 15 q^6 + 22 q^7 + 45 q^8 + 78 q^9 + 
 140 q^{10} + 238 q^{11} + 426 q^{12}+ \cdots$&$2,3, 4, 5, 6^4,  7^2, 8^6, 9^8, 10^8, 11^{10}, 12^9 ,\cdots$ \\
  \hline
  $4$ &$1 + q^2 + 2 q^3 + 5 q^4 + 7 q^5 + 18 q^6 + 29 q^7 + 63 q^8 + 
 112 q^{9} + 221 q^{10} + 400 q^{11} + 771 q^{12} + \cdots$ &$2,3,4^2, 5, 6^5, 7^4, 8^{11}, 9^{14}, 10^{24},11^{34}, 12^{55},\cdots $\\
  \hline
  $5$ & $1 + q^2 + 2 q^3 + 5 q^4 + 8 q^5 + 19 q^6 + 32 q^7 + 70 q^8 + 
 130 q^9 + 261 q^{10} + 490 q^{11} + 969 q^{12}+\cdots$ & $2, 3, 4^2, 5^2, 6^5, 7^5, 8^{13}, 9^{19}, 10^{34}, 11^{52}, 12^{94},\cdots$ \\
 \hline
  $6$ &$1 + q^2 + 2 q^3 + 5 q^4 + 8 q^5 + 20 q^6 + 33 q^7 + 73 q^8 + 
 137 q^9 + 279 q^{10} + 530 q^{11} + 1066 q^{12}+\cdots$&$2, 3, 4^2, 5^2, 6^6, 7^5, 8^{14}, 9^{21}, 10^{39}, 11^{62}, 12^{117}, \cdots$\\
 \hline
 $7$ & $1 + q^2 + 2 q^3 + 5 q^4 + 8 q^5 + 20 q^6 + 34 q^7 + 74 q^8 + 
 140 q^9 + 286 q^{10} + 548 q^{11} + 1106 q^{12} + \cdots$& $2, 3, 4^2, 5^2, 6^6, 7^6, 8^{14}, 9^{22}, 10^{41}, 11^{67} , 12^{127}, \cdots$ \\
  \hline
\hline
   \end{tabularx}
}
    \caption{The vacuum character for the vector coset model for fixed $k=3$ and small values of $N$.}
 \label{t:2}
  \end{center}
 \end{table}


To better understand the coset algebra in the infinite $k, \ell$ limit, it is instructive to find the algebra of the coset when $\ell$ is large but $k$ is fixed to a given value.  The vacuum character for such a coset is given by \equ{branching1} and the string functions continue to be given by \equ{stringfunction}. The string functions for a fixed level can be calculated in Mathematica using the package affine.m \cite{Nazarov:2011mv}. 

In Table \ref{t:2} we list the algebra for various $N$ for $k=3$. As can be seen, even for this low value of the level, the number of generators of the algebra grow quickly with the spin. 
In fact, for a fixed value of $N$, the coset algebra stabilizes for a small value of the level $k$ --- that is the coset generators do not change after a certain level. This fact was earlier reported in \cite{Blumenhagen:1991}. As we show in Table~\ref{t:3}, for $N=3$ the algebra has stabilized at level $8$. The field content at this level is identical to the field content at level $k=\infty$ calculated in \equ{n3char}. Note that the number of null states continue to change and specifically decrease as we increase the level to infinity. The null states, however, never disappear from the spectrum and are present even in the infinite level limit.

Note that the growth rate of currents at finite $k$ is sharper than what might expect from the $T$-dual coset $SU(k+\ell)_N/(SU(k)_N \times SU(\ell)_N )$. The symmetry algebra of this dual coset is expected to be a (truncation of) the matrix extension of $\W_N$ in the large $\ell, N$ limit for fixed $k$ --- so the multiplicity for a given spin would be at most $k^2$. However, it is possible for dual cosets to have different symmetry algebras \cite{Bowcock:1988}.

\begin{table}[!t]
 \begin{center}
\renewcommand{\arraystretch}{1.5}
\resizebox{0.995 \textwidth}{!}{
 \begin{tabular}{  c l c l}
 \hline
 \hline
 $k$ & Algebra& $k $ &Algebra\\
 \hline
 \hline
 $1$ &$2, 3.$ &$5$&$2,3, 4, 5, 6^4,  7^2, 8^7, 9^9, 10^{12}, 11^{16}, 12^{25},13^{25},14^{29},15^{27} $ \\
 \hline
 $2$ &$2, 3, 4, 5, 6^3, 7, 8^3,9^3$ &$6$&$2,3, 4, 5, 6^4,  7^2, 8^7, 9^9, 10^{12}, 11^{16}, 12^{26},13^{26},14^{32},15^{32},16^{8} $\\
 \hline
  $3$ &$2,3, 4, 5, 6^4,  7^2, 8^6, 9^8, 10^8, 11^{10}, 12^9 $&$7$&$2,3, 4, 5, 6^4,  7^2, 8^7, 9^9, 10^{12}, 11^{16}, 12^{26},13^{26},14^{33},15^{33},16^{11} $\\
  \hline
  $4$ &$2,3, 4, 5, 6^4,  7^2, 8^7, 9^9, 10^{11}, 11^{15}, 12^{22},13^{20},14^{16},15^{2} $&$8$ & $2,3, 4, 5, 6^4,  7^2, 8^7, 9^9, 10^{12}, 11^{16}, 12^{26},13^{26},14^{33},15^{33},16^{12} $\\
  \hline
  \hline
 \end{tabular}
}
 \caption{ \label{t:3}The algebra obtained from the vacuum character for the coset model $SU(3)_k/SU(3)$ for various $k$ values. The algebra stabilizes by $k=8$. }
  \end{center}
 \end{table}

 
\subsection{Asymptotic growth of vacuum character }  \label{asymptoticN}
The asymptotic growth of the vacuum character can be determined from the general formula for the asymptotic behaviour of branching functions \cite{Kac:1988}. 
Let us write the general branching function as
\be
	b_\l^\L(q) = \sum_s a_s q^s\,.
\ee
Then, asymptotically as $s\rightarrow \infty$,
\be
	a_s \sim \frac{1}{2} (c/6)^{1/4} b(\L,\l)  s^{-\frac{3}{4}} \exp\({\pi \sqrt{2/3\, c\, s}}\)\,
\ee
where, $ b(\L,\l)$ is a positive real number and $c$ is the central charge.
Thus, in our case:
\be
	a_s \sim s^{-\frac{3}{4}} \sqrt{N}\exp\( \pi\sqrt{{\tfrac{2}{3}N^2 s}}\)\,
\ee
where we have dropped the constants. 

In spite of the fact that there is an exponential increase in the number of currents, for small spin $s$, a large number of null states occur as $s$ becomes greater than $N^2$ for the vacuum branching function, as we saw in the previous section. Hence, the vacuum character has Cardy growth at large $s$. Note that this asymptotic behaviour holds only for finite $N$ and may change for infinite $N$.

\section{Generalized Casimir Current Algebra} \label{secCurrents}
In this section, we write down the explicit form of the currents for the stringy coset in \equ{coset2}. This coset and the associated current algebra have been studied in Refs.~\cite{Bouwknegt:1992wg,Bais:1987a}. We review the known facts about the current algebra for this coset at arbitrary level $k$ in \sct{sec:generalk}. We are interested in the current algebra in the limit of large level $k$. We show in \sct{sec:largek} that in this limit, the coset theory reduces to a theory of free bosons.  We also demonstrate that the number of currents grows with spin as expected from the vacuum character calculation in the previous section. We do this for the $N=2$ and $N=3$ cases in Secs.~\ref{sec:su2} and~\ref{sec:su3} respectively. Extrapolating from these results, we write down the general form of the current algebra generators in the large $N$ limit in \sct{sec:largeN}. We will identify the simplest of these generators  with the free field realization of the $\W^e_\infty[1]$ algebra.  The algebra $\W^e_\infty[\mu]$ is an infinite-dimensional sub-algebra of $\W_\infty[\mu]$, which consists of fields of even spin only \cite{Candu}. In \sct{sec:Relation}, we will show that a subset of the generators of our stringy coset can be arranged in representations of $\W^e_\infty[\mu]$. This is evocative of the higher spin square which is the symmetry algebra of the large $N$ symmetric product orbifold theory \cite{Gaberdiel:2015}. We will remark on this correspondence between the generators of the higher spin square and the coset theory in \sct{sec:Relation}. 

Here, we first establish background facts that we need to determine  the currents for the coset $SU(N)_k/SU(N)$.  For any coset algebra $G/H$, the generators are the currents of $G$ that commute with that of $H$. The generators of our coset algebra are composed from the $SU(N)_k$ generators: $J^a$, where $a$ varies from $1$ to $N^2-1$. These affine algebra generators, that transform in the adjoint representation of $SU(N)$, satisfy the following operator product expansion:
\be
\label{Jope}
	J^a(z)J^b(w) =  \frac{-k\,\delta_{ab}}{(z-w)^2} + \frac{f_{abc} J^c}{(z-w)} + \cdots\,.
\ee
Repeated indices will always imply summation, regardless of the placement of the indices.
The generators of the coset algebra are those operators of $SU(N)_k$ that commute with the $SU(N)$ currents, which are given by the zero modes of the affine currents. Thus, if $Q(z)$ is a generator of the coset algebra, then
\be
	\[J^0_{a}, Q(z)\]=0\,.
\ee
As is shown in Sec.~7.2.1 of Ref.~\cite{Bouwknegt:1992wg}, this implies that $Q(z)$ must be a differential polynomial invariant in the $SU(N)$ currents. The first such invariant is the stress-energy tensor
\be
	T(z)\sim \tfrac{1}{2} \no{J^a(z) J^b(z)}
\ee 
which is the quadratic Casimir of $SU(N)$ defined up to an overall normalization and $:\cdots:$ symbol denotes normal ordering. It is a quasi-primary field of conformal dimension two. We will find that the coset algebra currents, in general, take a simple form in the quasi-primary basis. A quasi-primary field is defined as having the following commutator with the Virasoro modes of the stress-energy tensor 
\be
\label{qpdef}
	\[L_m, Q_n(z)\] =\{ n- (d-1)m\}Q_{n+m}
\ee
where $m\in\{-1,0,1\}$. Here, $d$ is the conformal dimension of the field. A primary field on the other hand obeys \equ{qpdef} for all mode numbers $m$.

As is well-known, the Casimir invariants are independent symmetric polynomial invariants of $SU(N)$. In general, the number of polynomial {\it differential} invariants for a group, which is the set of possible currents for the coset CFT,  is much larger.  

\subsection{General $k$ and $k=1$} \label{sec:generalk}
At general level $k$ the stress energy tensor is given by
\be
\label{Tdef}
	T=\frac{-1}{2(k+N)}\no{J^a J^a}
\ee
The coset currents for spin $3$ and $4$ for general level $k$ have been written down in Ref.~{\cite{Bais:1987a}}, which we now review. 
At any level $k$ and for any $N$ there is always a single spin $3$ current of the form
\be
\label{Q3def}
	Q_3 =\alpha \, d_{abc} \no{J^a J^b J^c}\,.
\ee
Here, $\alpha$  is a normalization factor that is given up to a constant by
\be
	\alpha^2 = \frac{N}{(k+N)^2(N+2k)(N^2-4)}\,,
\ee
and $d_{abc}$ is the third-order invariant symmetric tensor for $SU(N)$. The above operator is, therefore, proportional to the third order Casimir of $SU(N)$. The normal ordering for three fields is defined as
\be
	\no{J^a J^b J^c} = \no{J^a   \no{J^b J^c}}
\ee
and in a similar manner for operators consisting of more fields.
Two primary spin $4$ currents were found for general $k$. The first field is
\be
\label{Q41}
	\frac{1}{4(k+N)} \big\{\no{ 2\partial^2 J^a  J^a}  - \,3 \no{\partial J^a \partial J^a }\big\} + \beta \no{TT} +~ \gamma \no{\partial^2 T}\,,
\ee
where $\beta$ and $\gamma$ are numerical factors dependent on $N$ and $k$.
The second field is given by
\be
\label{Q4}
	Q_4 = \alpha (k+N) d_{abcd} \no{J^a J^b J^c J^d} \,,
\ee
where $d_{abcd}$ is the fourth-order invariant symmetric traceless tensor of $SU(N)$. In general, for $SU(N)$, there are $N-1$ primitive $d$-tensors of order $2,\cdots,N$. Each of these corresponds to a current. As we will show below, however, there are other $SU(N)$ tensor invariants relevant to constructing coset currents.

The spin $4$ field  in \equ{Q4} occurs in the OPE of $Q_3(z)$ and $Q_3(w)$:
\begin{align}
\label{Q3ope}
	Q_3(z)Q_3(w) =& \frac{c/3}{(z-w)^6} +  \frac{2T(w)}{(z-w)^4} +  \frac{\partial T(w)}{(z-w)^3} \nn \\
				& +  \frac{1}{(z-w)^2} \bigg\{\frac{32}{22+5c} \no{T(w)T(w)} + \frac{3(c-2)}{10c+44}\partial^2 T(w) + Q_4(w) \bigg\} \nn\\
				&+ \frac{1}{z-w}  \bigg\{\frac{16}{22+5c}  \no{\partial(T(w)T(w))} + \frac{c-6}{15c+66}\partial^3 T(w) + \frac{1}{2} \partial Q_4(w) \bigg\} + \cdots
\end{align}

A well-known result is that for the coset  $SU(N)_1/SU(N)$, primary fields with spin higher than $N$ are either null or vanish and that there is only a single field at a given spin, so that the algebra becomes identical to the $W_N$ algebra. Let us show this for $N=3$, for the spin $4$ operators. The field in \equ{Q4} vanishes for $SU(3)$ as the tensor $d_{abcd}$  collapses to zero for $N<4$.  The second spin $4$ field in \equ{Q41} can be written as
\be
\label{Q34}
	Q^{N=3}_{4}~=~ \no{TT}- ~\frac{33+ 31k}{12(k+3)^2}\no{ \partial J^a \partial J^a }- ~\frac{6+4k}{9+3k}\, \partial^2 T \,.
\ee
While in Ref.~{\cite{Bais:1987a}} it was shown that the field in \equ{Q34} vanishes upon using an explicit realization of the Kac-Moody algebra in terms of free boson vertex operators (so that the Sugwara construction for $\W_N$ maps to the free field Miura realization), we can also directly compute the two-point function for $Q^{N=3}_{4}(z)$. This is given by
\be
	\frac{k(-99-60k+97 k^2 + 62 k^3)}{(3+k)^4}\,,
\ee
omitting overall numerical coefficients. The only non-zero integer value for which this vanishes and the primary field $Q^{N=3}_{4}(z)$ becomes null is $k=1$.


\subsection{The limit $k\rightarrow \infty$} \label{sec:largek}

In this section, we will construct the coset currents in the large $k$ limit, which is the main objective of this paper. We will work at finite $N$ and then extrapolate our results to large $N$. The coset currents for finite $k$ explicitly depend on $k$. To remove this dependence, we will redefine the $SU(N)_k$ generators as follows
\be
	J^a \rightarrow \frac{1}{\sqrt{k} } J^a \,.
\ee
As a result of this redefinition the stress-energy tensor in \equ{Tdef} becomes
\be
\label{Tinfdef}
	T=- \frac{1}{2} \,J^a J^a
\ee
in terms of the new generators. Similarly the spin $3$ current in \equ{Q3def} (and other higher spin currents) become $k$-independent under this redefinition. The OPE in \equ{Jope} between the Kac-Moody currents becomes
\be
	J^a(z)J^b(w) =  \frac{-\,\delta_{ab}}{(z-w)^2} \,,
\ee 
since the single-pole term is now suppressed by $\sqrt{k}$. The currents therefore, become essentially free in the large $k$ limit and the theory behaves like a theory of $N^2-1$ free bosons. 

To maintain continuity with finite $k$, we will continue to take the coset currents to be $SU(N)$ invariants. 
The behavior of the theory in the large $k$ limit can also be motivated as follows. The $SU(N)_k$ current algebra can be written in terms of $N-1$ bosons and so-called ``generating'' parafermions \cite{Gepner:1987}. The bosonic fields are denoted by $\phi_i$, where $1\leq i \leq N-1$ and expressed as a vector $\boldsymbol{\phi}$.
The parafermions $\psi_{\a}$ are fractional spin fields associated with the root lattice of $SU(N)$. Thus, here $\a$ labels the roots of $SU(N)$. The conformal dimension of these fields is given by
\be
	\Delta(\psi_\a) = 1- \frac{\a^2 }{2 k} = 1- \frac{1}{k}\,.
\ee
where we have used the normalization $\a^2=2$. 
In terms of these parafermions $\psi_{\a}$ and the bosonic field $\boldsymbol{\phi}$ the $SU(N)_k$ generators take the form
\be
\label{ve1}
	J_i(z) \sim \sqrt{k}\, \a_i  \cdot \partial_z \boldsymbol{\phi}
\ee
when $J_i(z) $ belongs to the Cartan sub-algebra and the form
\be
\label{ve2}
	J_\a (z) \sim \sqrt{k} \psi_{\a}  \exp[i  \a \cdot \boldsymbol \phi(z)/\sqrt{k}]
\ee
for the rest of the generators corresponding to the $N^2-N$ roots of $SU(N)$. For level $k=1$, the parafermions $\psi_{\a}$ have vanishing dimension and the generators reduce to the usual vertex operator representation of the current algebra in terms of $N-1$ bosonic fields. As $k\rightarrow \infty$, the parafermions are promoted to bosons (of spin one). The $\exp$ term in \equ{ve2} reduces to one and the form of the generators $J_{\a}(z)$ become similar to $J_i(z)$. This is often referred to in the literature \cite{Bakas:1990} as flattening of the $SU(N)_k$ algebra in the large level limit to a $U(1)^{N^2-1}$  algebra. 

In the following, we will write down the generators of the coset algebra in the infinite level limit for the cases $N=2$ and $N=3$. We will denote a quasi-primary field of spin $s$ by $Q_s$ and a primary field by $P_s$. The associated $N$ value should be clear from context. We will always define the fields up to an overall normalization. To find the quasi-primary and primary currents, we have used the Mathematica package OPEdefs \cite{Thielemans:1991}.

\subsubsection{$N=2$} \label{sec:su2}
The $N=2$ case was studied in Ref.~\cite{deBoer:1993}. A set of classical currents for the stringy $SU(2)$ coset can be obtained by acting on the Casimir invariant $\Tr(J J)$ by derivatives. These currents are of the form
\be
\label{bilinear}
	\Tr\big(\partial^\mu J \partial^\nu J\big)\,.
\ee
In addition to the Casimir invariant, $SU(2)$ has cubic invariants given by
\be
\label{trilinear}
	\Tr([\partial^\mu J ,\partial^\nu J]\partial^\gamma J) \;\;\; \forall \mu \neq \nu \neq \gamma\,.
\ee

We can count the number of these invariants. The number of bilinear terms is the number of ways one can divide an integer into exactly two parts. The number of trilinear terms is given by the generating series for the number of ways to divide an integer into three  {\it distinct} parts. The number of ways to divide an integer into $p$ distinct parts is given by the generating function:
\be\label{kdistinct}
	\frac{q^{p(p+1)/2}}{(1-q)(1-q^2)\cdots(1-q^p)}\,.
\ee
To get  the independent terms we remove the total derivatives.
Then the generating function for the classical currents is given by
\be
	\frac{q^2}{(1-q^2)} + \frac{q^6}{(1-q^2)(1-q^3)}= q^2+ q^4+ 2 q^6 + 2 q^8 + q^9 + 2 q^{10} + q^{11} + 2 q^{12} + \cdots\,.
\ee
To find the quantum algebra of the coset, we have to find the primary completion of the classical currents. Relations between the invariant tensors of $SU(2)$ will make some of these currents vanish. This together with the presence of null states will truncate the set of infinite currents to the finite set listed in \equ{SU2currents}, derived from the vacuum character. In fact, it can be shown that the first vanishing state arises at the order at which a syzygy (a relation between the invariants) is present. For more details regarding this, the reader is referred to \cite{deBoer:1993}. 

We now write down the explicit form of the currents up to an overall normalization. The stress energy tensor of the coset is given by \equ{Tinfdef}. There is no spin $3$ current.  We have a single primary field of dimension $4$ which is of the form 
\be
\label{n2p4}
	P_4 =  \big(\tfrac{1}{2}:\partial^2 J^a J^a: -~ \tfrac{3}{4}:\partial J^a \partial J^a: \big)- \tfrac{9}{37} :T T: + \tfrac{30}{37} \partial^2 T \,.
\ee
Note that the first term in brackets is a quasi-primary field of dimension $4$ while the rest  are correction terms to make the field primary.  Next we have two primary fields of spin $6$. The first is of the form in \equ{bilinear} and is given by
\begin{align}
\label{n2p61}
	P_{6,1} = \no{\partial^3 J^a \partial J^a} &-\no{ \partial^2 J^a \partial^2 J^a} - \tfrac{1}{10}\no{\partial^4 J^a  J^a}\  \nn \\
&+ ~\alpha \no{TTT} + ~\beta\no{\partial^2T T} +~ \gamma \no{\partial T \partial T}  +~ \delta  \no{Q_4 T} +~\epsilon \,\partial^2 Q_4+ \zeta\, \partial^4 T \,.
\end{align}
The term in the first line of the RHS is the associated quasi-primary field. The coefficients $\a,\b,\cdots$ are given in Appendix~B.
The second field is of the form \equ{trilinear} and is given by
\be
\label{prim3}
	P_{6,2}~=~\epsilon_{abc} \no{J^a \partial J^b \partial^2 J^c}\,
\ee
where $\epsilon_{abc}$ is the Levi-Civita tensor. The expression as written above is already a primary field of dimension $6$ and does not need any correction terms. Continuing in this manner, one can write down all operators of the algebra. As stated above, the primary fields start becoming null from spin $11$ and the algebra thus consists of a finite number of fields. 

\subsubsection{$N=3$} \label{sec:su3}
For the stringy $SU(3)$ coset, the classical currents related to the Casimir invariants are of the form
\begin{align}
\label{su3casimirs}
	&\Tr(\partial^\mu J\partial^\nu J) \equiv  \partial^\mu J^a \partial^\nu J^a \,,\nn\\
	&\Tr(\{\partial^\mu J, \partial^\nu J\} \partial^\gamma J) \equiv d_{abc} \partial^\mu J^a \partial^\nu J^b \partial^\gamma J^c \,.
\end{align}
Below we list all the currents up to spin $6$. The lowest spin operators having the structure in \equ{su3casimirs} are the stress-energy tensor and the spin $3$ Casimir current:
\be
	P_3 = d_{abc} \no{J^aJ^b J^c}\,. 
\ee
The lowest spin currents with additional derivatives are the spin $4$ current of the form 
\be
\label{n3q4}
	Q_4= \tfrac{1}{2}\no{ \partial^2 J^a  J^a}  - \,\tfrac{3}{4} \no{\partial J^a \partial J^a }
\ee
and the spin $5$ current of the form 
\be
	Q_5 = d_{abc}\, \big( \no{\partial^2 J^a J^b  J^c}  -\,\tfrac{3}{2} \no{\partial J^a \partial J^b  J^c} \big) \,.
\ee 
At the next level there are two additional spin $6$ currents of the form 
\be
\label{b6}
	Q_{6,1} =\no{\partial^3 J^a \partial J^a} -\no{ \partial^2 J^a \partial^2 J^a} - \tfrac{1}{10}\no{\partial^4 J^a  J^a} 
\ee
and
\be
\label{s6}
	Q_{6,2} = d_{abc}\, \big( \no{\partial^3 J^a  J^b  J^c} -\,6 \no{\partial^2 J^a \partial J^b  J^c} +\, 6\no{ \partial J^a \partial J^b \partial J^c} \big) .
\ee 
The currents as written are all quasi-primary, save for the spin $3$ current $P_3$ which is primary. The primary completion of these fields appear in Appendix~B.

Apart from the Casimir invariants, there are other invariants as well for $SU(3)$.   The tensor $f_{abc}$ is a skewsymmetric invariant and it leads to the currents of the form
\be
	 f_{abc}\, \partial^\mu J^a \partial^\nu J^b \partial^\gamma J^c~~ \forall ~~\mu \neq \nu \neq \gamma.
\ee
Indeed the first such current is
\be
\label{as6}
	P_{6,3} = f_{abc}\no{\partial^2 J^a \partial J^b  J^c}\,,
\ee
which is a primary field of dimension $6$. This accounts for three of the four spin $6$ currents predicted by the vacuum character.  

To write the final spin $6$ current it is useful to look at the primary fields in the theory. Primary fields of the coset CFT can be divided into two categories: those that are $SU(3)$ singlets and those that are not. The vacuum character contains information about primary fields that are also $SU(3)$ singlets. Taking the vacuum character in \equ{n3char} and expanding in Virasoro characters gives
\be
\label{VirExp}
	(1-q)V_0(q) + V_2(q) + V_3(q) + V_4(q) +  V_5(q) + 5 V_6(q) + 3 V_7(q) + 11 V_8(q) + \cdots. 
\ee
where $V_h(q)$ is the character of the Virasoro algebra Verma module 
\be
	V_h(q) = q^h \prod_{j=1}^{\infty} \frac{1}{1-q^j}\,.
\ee
and we have assumed that the Virasoro characters are irreducible. From \equ{VirExp}, we can see that there should be a total of five primary fields of spin $6$, that are also $SU(3)$ singlets. These include the fields that are composite operators. Composite operators can also be divided into two categories: those that are composed of $SU(3)$ singlets and those that are composed of fields transforming non-trivially under $SU(3)$. An example of the first kind of operator is the following composite spin $6$ operator:
\be
	P_{6,4}=\,\no{P_3 P_3}+\, \alpha \no{TTT} + \beta \no{Q_4 T}+ \, \gamma\no{\partial^2 T T} + \, \delta \no{\partial T \partial T} + \, \epsilon \, \partial^2 Q_4 + \, \zeta\, \partial^4 T.
\ee
This is a primary field if the coefficients $\a,\b,\gamma,\delta,\epsilon,\zeta$ take the values
\be
	\a= \frac{503}{372}, ~\b= -\frac{33}{40}  , ~\gamma=\frac{2627}{1240}, ~\delta=\frac{413}{124}, ~\epsilon=\frac{37}{80}, ~\zeta=-\frac{101}{310}\,.
\ee

To find the other $SU(3)$ invariants we look for primary fields that are not $SU(3)$ singlets. A spin $2$ field of this nature was introduced in Ref.~\cite{Bais:1987a}, and takes the form
\be
	P^a_2 = d^{a}_{\phantom{a}bc}\no{J^b J^c}\,.
\ee
This is not the only possible primary field transforming in the $SU(3)$ adjoint rep. Fields of the schematic form $d_{abc} \no{\partial^\mu J^a \partial^\nu J^b}$ can be potential primaries. The operator
\be
	P^a_4 = d^{a}_{\phantom{a}bc}\( \no{\partial^2 J^b J^c} - \tfrac{3}{2}  \no{\partial J^b \partial J^c}\) + \tfrac{3}{14} \no{\partial^2 P^a_2} - \tfrac{5}{7} \no{P^a_2 T} \,.
\ee
is a (non-null) spin $4$ primary. In fact, we can generate primaries from the skew-symmetric tensor invariant $f_{abc}$ in the same manner. The field
\be
	P^a_3 = f^{a}_{\phantom{a}bc}\no{\partial J^b J^c}\,
\ee
is a primary operator of dimension three. 

We can construct new $SU(3)$ singlets from such primary fields. It will not, however, always be the case that the operators generated in this way are distinct to the singlets already constructed or are not null. The composite spin $4$ primary 
\be
	\no{P^a_2 P^a_2} -\tfrac{40}{31} \partial^2 T -\tfrac{220}{93} \no{T T}
\ee
is the same as the primary completion of the spin $4$ field in \equ{n3q4}. However, the composite field
\be
\label{p65}
	P_{6,5} =\no{P^a_2 P^a_4} +\tfrac{15}{7}\no{T T T} - \tfrac{17}{42}\no{\partial^2 T T} +\tfrac{5}{21} \no{\partial T \partial T} +\tfrac{31}{14} \no{Q_4 T} + \tfrac{62}{63} \partial^4 T - \tfrac{527}{252}\partial^2 Q_4
\ee
is a new spin $6$ primary field. It can be easily verified, using Mathematica,  that the operators $P_{6,1}, P_{6,2}, P_{6,3}, P_{6,4}$ and $P_{6,5}$ are linearly independent. Note that we can construct another spin $6$ singlet of the form $\no{P_3^a P_3^a}$,  but this operator is a linear combination of $P_{6,1}, P_{6,2}$ and $P_{6,5}$.

The primary field in \equ{p65} is a composite field as far as the CFT is concerned, but is an independent $SU(N)$ invariant. It will, therefore, be counted by the vacuum character of the coset CFT. The spin $6$ fields that contribute to the vacuum character are thus: $P_{6,1}, P_{6,2}, P_{6,3}$ and $P_{6,5}$.  Note that the field $P_{6,4}$ is a ``double-trace'' $SU(N)$ invariant and its contribution to the vacuum character has already been accounted for by $P_3$.  The number of independent currents at spin $6$ is thus four as predicted by the vacuum character. We see that unlike for the $N=2$ case, the $N=3$ stringy coset also needs composite currents to generate the full set of currents. Of course, the currents $T,P_4$ etc are also composites of the primary field $J^a$, but we use the word composite here to mean operators that are composites of primary fields that are themselves composite in $J^a$.  It is not difficult to estimate the number of such composite currents, although it is hard to discern which of them are non-redundant \cite{Dittner:1972}. For $SU(3)$, invariant tensors that can be formed out of the composite operators, denoted here by $A^a,B^a,C^a$, take the form
\be
	d_{ab} A^a B^b\,, ~~~~d_{abc}A^a B^b C^c\,.
\ee
The growth rate for such composite operators (with increasing spin) exceeds the growth rate for generators predicted by the vacuum character in \equ{n3char}. However, it has to be checked on a case-by-case basis which of these fields are independent and contribute to the vacuum character.

All the generators that we have constructed for infinite $k$ are also present at finite $k$ (for $k\geq3$). It is worthwhile to write down the exact form of some of these generators for finite $k$. At finite $k$, the spin $4$ primary field takes the form
\be
	P_4(k)= Q_4 - \tfrac{9 (3 + k)^2}{33 + 31 k}  \partial T + \tfrac{ 30 (3 + k)^2} {33 + 31 k} \no{T T}\,.
\ee
The bilinear spin $6$ primary current takes the form
\begin{align}
	P_{6,1}(k)= Q_{6,1} &- \tfrac{21 \a ^2(-129+106k +55k^2)}{5 \b\g\delta} \no{\partial^2T T} - \tfrac{42 \a^2(1+k)(6+k)}{  \b\g\delta}  \no{\partial T \partial T} + \tfrac{21 \a }{5\b} \no{Q_4 T}  + \nn \\ &+ \tfrac{2 \a^2 (-219+  1126 k +790 k^2) }{5  \b\g\delta}\partial^4 T - \tfrac{49 \a }{10\b} \partial^2 Q_4 - \tfrac{42 \a^3 (3-7k)}{\b\g\delta}\no{T T T}\,
\end{align}
where $\a= 3+k, \b= 9+4k, \g= 1+5k, \delta= 51+31 k$\,. In the above equations, the stress-energy tensor is defined as in  \equ{Tdef} and the the currents $J^a$ are rescaled as $\sqrt{k}J^a$. The form of the quasi-primary operators is thus independent of $k$ for the bilinear currents. This is not true for the generator $P_{6,3}$. This current is not primary at finite $k$ and has to be modified in the following way to stay primary
\begin{align}
	P_{6,3}(k)=& f_{abc} \no{\partial^2 J^a \partial J^b J^c}  -\tfrac{21 \a^2(1353+6518k +2225k^2)}{5 \b\g\delta} \no{\partial^2T T} - \tfrac{42 \a^2 (159+160k+61k^2)}{5  \b\g\delta}  \no{\partial T \partial T} +   \nn \\& + \tfrac{23 \a }{5\b} \no{Q_4 T}  + \tfrac{2 \a (-2529 +8379k +9668 k^2 +2150 k^3) }{5  \b\g\delta}\partial^4 T  + \tfrac{474 +157k}{30\b} \partial^2 Q_4 - \tfrac{2 \a^3 (3+193k)}{\b\g\delta}\no{T T T}\,.
\end{align}
Note that as $k \rightarrow \infty$ the fields $P_4(k)$ and $P_{6,1}(k)$ reduce to their corresponding counterparts in \equ{An3p4} and \equ{An3p61}, while $P_{6,3}(k)$ becomes identical to \equ{as6}.

\subsubsection{Large $N$} \label{sec:largeN}
\begin{figure}[t!] 
\begin{center}
\includegraphics[scale=1]{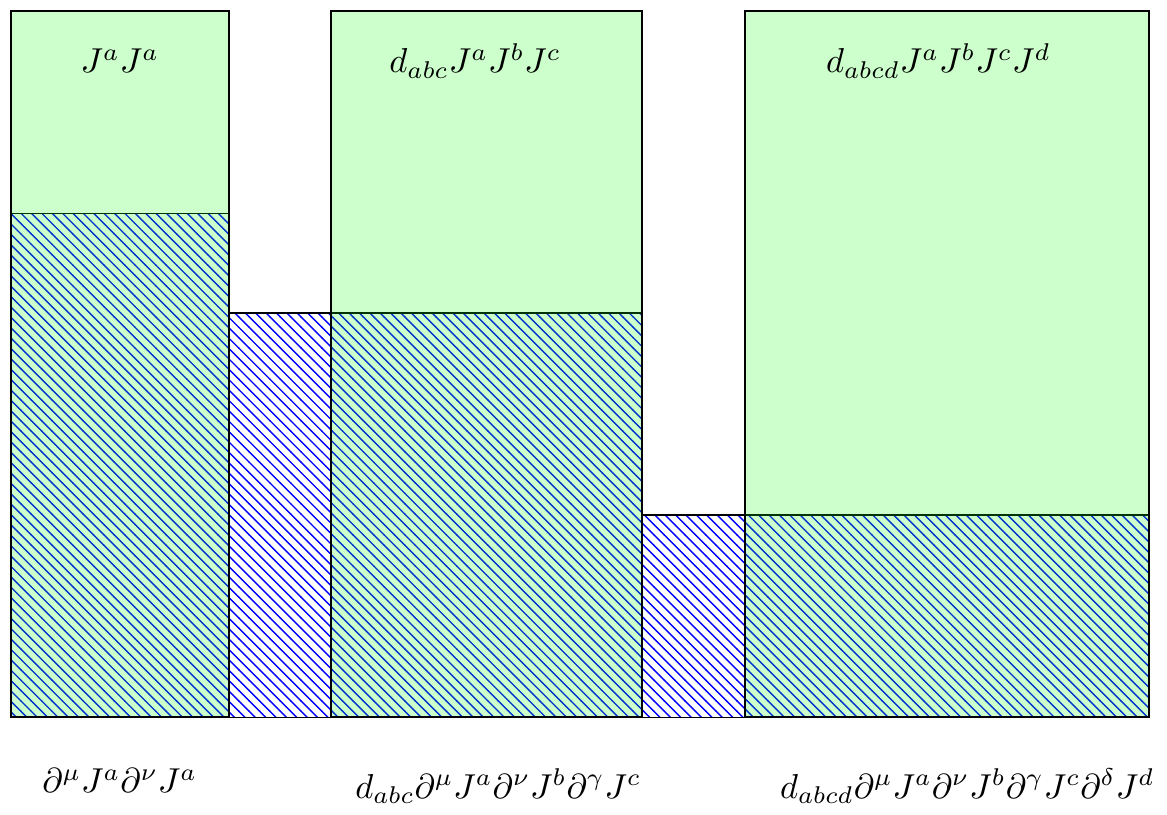}
\caption{\label{rectfig} The operators in the top-most row correspond to the $SU(N)$ Casimir invariants. A subset of generators for the stringy coset algebra is generated by acting by derivatives on the constituent terms of each Casimir operator. The cross-hatched area denotes the operators  that are null and not part of the algebra for a given level $k$ and $N$. For $k=1$, only the top row of operators is not null and corresponds to the $W_N$ algebra.  Increasing the level $k$ reduces the cross-hatched area but does not eliminate it completely even as $k$ tends to $\infty$. Increasing $N$ corresponds to adding more columns and also reducing the cross-hatched area.  As we take $N$ to $\infty$, with $k$ already taken to $\infty$, null states disappear from the first column and the algebra becomes $\W_\infty^e[1]$. }
\label{figrect}
\end{center}
\end{figure}
At large $N$, the currents that follow from the symmetric $d$-tensor invariants are given by
\be
\label{genericC}
	d_{abc\cdots}\partial^\m J^a \partial^\n J^b \partial^\g J^c \cdots\,.
\ee
For $SU(N)$ there is a single Casimir invariant at orders $2,\cdots,N$. 
Hence for infinite $N$, the generating function for the generalized Casimir currents is
\be
\label{SCF}
	\sum_{p=2}^{\infty} \frac{q^{p}}{\prod_{k=2}^{p}(1-q^k)} = \prod_{n=2}^{\infty} \frac{1}{(1-q^n)} -q\,.
\ee
Since, the tensors $d_{abc\cdots}$ are totally symmetric, the generating function for an order $p$ current of the form in \equ{genericC} is given by the number of ways to divide an integer into $p$ parts.
Classically, thus, the number of independent currents grows at least as fast as $\exp({\sqrt{n})}$. If one assumes that at large $N$, there are no null fields in this set of currents, then the growth of this set of currents matches that of the higher spin square algebra. 

Clearly as we saw from the above examples of $SU(2)$ and $SU(3)$, these are not the only possible currents. For the group $SU(N)$, there are skew-symmetric invariant tensors $f_{abc\cdots}$ of order $3,5,\cdots,2N-1$ which lead to antisymmetric currents. In the literature, these skew-symmetric tensors are also known as $\Omega$ tensors \cite{deAzcarraga:1997, deAzcarraga:2000}. The total number of such states as $N \rightarrow \infty$, assuming no null states is given by
\be
\label{ASCF}
	  \sum_{\substack{k=3 \\  k \in \textrm{odd}}}^{\infty} \frac{q^{k(k+1)/2}}{(1-q^2)\cdots(1-q^k)} = \frac{1-q}{2} \bigg \{\prod_{n=1}^{\infty} (1+q^n) - \prod_{n=1}^{\infty} (1-q^n) \bigg \} -q\,.
\ee

As we saw for finite $N$, composite currents are also present for the coset CFT. In general, it is hard to count these currents.  Invariants composed of primary fields that transform non-trivially under $SU(N)$ do not always lead to new or non-null currents. This happens because of the identities that exist between $SU(N)$ tensors (see, for example, Appendix~B of \cite{Bais:1987a}).

\subsection{Relation with $\W_\infty^e[1]$ and the higher spin square} \label{sec:Relation}
We now focus on the currents bilinear in the $J^i$'s. As we saw for finite $N$, the independent bilinear currents at any value of the level $k$  have even spin only with multiplicity one at each spin. In the quasi-primary basis, these currents do not change with increasing $k$ or $N$, since only the overall normalization changes. (In the primary basis, this is no longer true). In the large level limit and at finite $N$,  the coset theory reduces to that of $N^2-1$  free bosons, hence the bilinear currents can be identified with a realization of the $\W_\infty^e[1]$ algebra at central charge $N^2-1$. As is well known, such a realization is a finite truncation of the $\W_\infty^e[1]$ algebra \cite{Blumenhagen:1994}. As we take $N\rightarrow \infty$, we recover the full $\W_\infty^e[1]$ algebra. 

It is natural to ask whether the higher-order generators of the stringy coset algebra as arranged in Fig.~(\ref{figrect}) can be identified with representations of the $\W_\infty^e[1]$ algebra. The OPE of any  $\W_\infty^e[1]$  operator, which are the bilinear operators, with a generator of order $p$ gives rise to operators of the same order. For example, the OPE of any bilinear term of the form $\no{\partial^\mu J^a \partial^\nu J^a}$ with a generic trilinear term is
\begin{align}
		&\no{\partial^\mu J^a \partial^\nu J^a (z) }\no{d_{bcd} \partial^\a J^b \partial^\b J^c \partial^\gamma J^d(w)} \nn\\
		\sim\, &\frac{\delta^{ab} \delta^{ac} d_{bcd} \partial^\gamma J^d(w)}{(z-w)^{\mu+\nu+\a+\b+4}}~+~\frac{\delta^a_{b} d_{bcd} \no{\partial^\nu J^a(z)  \partial^\b J^c(w) \partial^\gamma J^d(w)}}{(z-w)^{\mu+\a+2}} \nn\\
		\sim\, &\frac{ d_{bcd}  \no{\partial^\nu J^b(w)  \partial^\b J^c(w) \partial^\gamma J^d(w)}}{(z-w)^{\mu+\a+2}} + \frac{ d_{bcd} \no{ \partial^{\nu+1} J^b(w)  \partial^\b J^c(w) \partial^\gamma J^d(w)}}{(z-w)^{\mu+\a+1}}  +\cdots
\end{align}
We have written the OPE schematically omitting numerical coefficients in all terms and writing only a {\it single} representative term for different possible ways of contracting. The first term in the second line vanishes because the tensor $d_{bcd}$ is traceless. Thus we are only left with trilinear operators in the OPE. The same logic applies to the OPE of any $p$-th order operator with a bilinear operator. The operators in each column of  Fig.~(\ref{figrect}) thus fall into a representation of $\W_\infty^e[1]$ . 

Let us identify the $\W_\infty^e[1]$ representation that corresponds to each column of operators in Fig.~(\ref{figrect}). We use the standard coset notation for representations of the $\W_\infty^e[1]$ alegbra. Conventional $\W$-algebras that we deal with in this paper are the symmetry algebras of cosets of the form $\frac{\mathfrak{g}_{k} \otimes \mathfrak{g}_{1}}{\mathfrak{g}_{k+1}}$. The  notation $(\Lambda_+; \Lambda_-)$ is used to denote a representation of a $\W$-algebra where $\Lambda_+$ is a representation of $\mathfrak{g}_{k}$ and $\Lambda_-$ is a representation of $\mathfrak{g}_{k+1}$. Then, a order $p$  column corresponds to the representation ${([0^{p-1},1,0,\ldots,0];0)}$ of $\W_\infty^e[1]$.  The wedge character of a representation $R$ of $\W_\infty^e[\mu]$ is given by 
\be
	 q^{\frac{\mu}{2} B(R^T) }\chi^{U(\infty)}_{R^T}\,,
\ee
where $R^T$ is the transpose of the representation $R$, $B(R^T)$ is the number of boxes in the Young tableaux of $R^T$ and $\chi^{U(\infty)}$ is the associated Schur function (See  \cite{Gaberdiel:2011, Gaberdiel:2015}  for details). 
Thus the wedge character of the representation ${([0^{p-1},1,0,\ldots,0];0)}$ is given by
\be \label{bwedge1}
	b^{({\rm wedge}) [\lambda=1]}_{([0^{p-1},1,0,\ldots,0];0)} = \frac{q^{p/2} q^{p/2}}{\prod_{k=1}^{p}(1-q^k)}\,
\ee
which is the same as the generating function of a column of operators with order $p$. Note that these are the same representations that constitute the higher spin square, the algebra of the large $N$ symmetric product orbifold.

The operators of the coset algebra look very different to corresponding operators of the higher spin square algebra. Nevertheless, the number of operators (and hence the character of the representation) is the same in a column of the coset algebra whose highest weight state is an operator of the form $d_{ab..}J^a J^b ..$ of order $p$ and in a column of the higher spin square (See Fig.~\ref{squafig}) whose highest weight state is of the form $J^a J^a..$ of the same order $p$. Identical characters imply identical representations, since a representation of an algebra has a unique character. Thus, the subset of generators of the coset algebra that are present in Fig.~(\ref{figrect}) must be isomorphic to the higher spin square. 

Generators of the higher spin square can also be organized in terms of a $\W_{\infty}[0]$ algebra which is called the horizontal sub-algebra in Ref.~\cite{Gaberdiel:2015}. In the coset case, this means that there should exist a change of basis for the $SU(N)$ currents, such that in the new basis the generators in the top row of Fig.~(\ref{figrect}) close at infinite $k$. It is possible that such a basis exists, since the coset theory also has a free fermion formulation at $c=N^2-1$. As is well known, the $\W_{\infty}[0]$ algebra can be expressed in terms of free fermions \cite{Pope:1991}.

Next we look at operators of the generic form $\no{f_{bcd\cdots} \partial^\a J^b \partial^\b J^c \partial^\gamma J^d(w)\cdots}$. The generating function for this set of operators is given by \equ{kdistinct}. Interpreting this as a wedge character of $\W_\infty^e[1]$, we find 
\be
		\frac{q^{p(p+1)/2}}{{\prod_{k=1}^{p}(1-q^k)}}= \frac{q^{p/2} q^{{p^2}/2}}{{\prod_{k=1}^{p}(1-q^k)}} = b^{({\rm wedge}) [\lambda=1]}_{([p,0,\ldots,0];0)}   \,.
\ee
that it corresponds to the representation ${([p,0,\ldots,0];0)}$. All the ``elementary operators'' of the stringy coset algebra can thus be organized into representations of $\W_\infty^e[1]$.

\section{Discussion}
In this paper, we have examined the coset in \equ{gencoset} in the free field limit, which is equivalent to its zero coupling limit. On the basis that the central charge of this coset scales as $N^2$, it has generally been expected that this coset is dual to a string theory in the bulk. We computed the vacuum character for the coset at finite $N$ and found that the currents of the symmetry algebra indeed exhibit an exponential growth with the spin. The $\W$-algebra of the coset in \equ{gencoset} in the free field limit is the same as the $\W$-algebra of the coset in \equ{coset2} in the large $k$ limit. We have written down the explicit form of the low-dimension currents for the second coset model.  As we saw in the text, the currents of this coset are simply $SU(N)$ invariants composed out of differential operators of the form $\partial^\mu J^a$, where the $J^a$ are $SU(N)_k$ currents. As such, this coset theory is an exact analog in two dimensions of $SU(N)$ Yang-Mills theories in higher dimensions, with the  addition of Virasoro symmetry.

The zero coupling limit of supersymmetric Yang-Mills theory in four dimensions is expected to be dual to tensionless string theory on the $AdS_5$ background \cite{Sundborg:2000}. It is, therefore, of interest to ask what the bulk dual of our coset model in the free field limit is. We know, from considerations of the D1-D5 system in Type IIB string theory, that it is a symmetric product orbifold theory that is expected to be dual to string theory on the $AdS_3$ background. Indeed, there has been much recent work in this direction clarifying aspects of this duality between tensionless strings and the free symmetric product orbifold theory \cite{Eberhardt:2018}. It is not clear where the coset theories, we have considered in this paper, fit into this picture. A more detailed understanding of the moduli space of $AdS_3$ string theories would be useful \cite{OhlssonSax:2018}. Further hints may come from integrability \cite{Sax:2014}.

How does the more general $\W$-algebra of the stringy coset models relate to the $\W_\infty[\mu]$ symmetry of the vector coset models? We found that in the free field limit, the algebra $\W^e_\infty[1]$ is a sub-algebra of the full coset algebra. Further, there is a distinguished set of generators of the coset algebra that can be arranged into representations of $\W^e_\infty[1]$. Operators that are directly derived from the symmetric tensor invariants of $SU(N)$ can be arranged in the ${([0^{p-1},1,0,\ldots,0];0)}$ representations of $\W^e_\infty[1]$. Operators that are related to antisymmetric invariants of $SU(N)$  can be arranged in the ${([p,0,\ldots,0];0)}$ representations of $\W^e_\infty[1]$. Since the first set of operators can be organized in the same set of representations of $\W^e_\infty[1]$ as the operators of the higher spin square, we propose that this set of operators is identical to the higher spin square.
The higher spin square also has a $\W_{\infty}[0]$ horizontal algebra, in addition to the vertical $\W^e_\infty[1]$ algebra. We have not explicitly identified this horizontal algebra in the coset case. It is important to do so, in order to cement the identification of the coset sub-algebra with the higher spin square.
 
In addition to the ``elementary'' generators, the coset theory also has a large number of composite operators at general values of $N$. In this paper, we have not attempted to classify them in representations of $\W^e_\infty[1]$. It is obviously of interest to understand the nature of these generators to comprehend the full symmetry algebra of the stringy coset theory.  The coset in \equ{gencoset}  is $T$-dual to a coset which is holographically dual to Vasiliev theory with a matrix extension. It would be interesting to explore the exact relation between this stringy coset and the matrix cosets \cite{Creutzig:2018}.
In this paper, we have worked in the limit of zero coupling. However, in general the coset algebra depends on the parameter $\lambda$. It would be interesting to find how the algebra changes once the coupling is switched on. At certain values of non-zero $\lambda$, for example at $\lambda=1$, the coset theory has a formulation in terms of free bosons/fermions. This is also the point, where the symmetry algebra of the coset enhances to an $\N=1$ supersymmetric algebra. It would be nice to do an analysis similar to this paper for the $\lambda=1$ theory. More generally, Wolf space coset generalizations of \equ{gencoset} can be studied in a similar manner.

\section*{Acknowledgements}
We thank Rajesh Gopakumar, Yang-Hui He, Bogdan Stefanski and Alessandro Torrielli for discussions. We thank Biswajit Ransingh for running a computer program for us at HRI, Allahabad.
\appendix 
\section{The $SU(N)$ tensor invariants}
The tensors $d_{abcd..}$ are totally symmetric tensors which we have also chosen to be traceless. Suitable traceless symmetric tensors are defined in \cite{deAzcarraga:1997}, where they are referred to as $t$-tensors.

We use the notation $\d$ for the standard symmetric invariant  tensors of $SU(N)$. They are
defined recursively \cite{Sudbery:1990} starting from the standard third-order
symmetric tensor $\d_{ijk}$ . One can construct the tensor
\be 
\label{recursion} \mathfrak{d}^{(r+1)}{}_{{i_1} \, \dots \, {i_{r+1}}}= 
\d^{(r)}{}_{{i_1} \, \dots \, {i_{r-1}}j} \, \d^{(3)}{}_{j{i_r}
	{i_{r+1}}} \quad , \quad r=3,4, \dots \quad . 
\ee \noindent
For $r \geq 3$, the above construction does not define totally symmetric 
tensors. The $\d$-family of symmetric tensors is obtained by 
symmetrising over all free indices in (\ref{recursion}).

The $SU(n)$ $d$ tensors are related to members of the $\d$-family in the following way
\begin{align} \label{tdefs} 
d{}_{ij} & \sim \delta_{ij} \,, \nn \\
d{}_{ijk} & \sim  \d_{ijk} \,, \nn \\
d{}_{ijkl} & \sim n(n^2+1) \d^{(4)}{}_{(ijkl)} -2(n^2-4) \delta_{(ij}\delta_{kl)} \,, \nn  \\
d{}_{ijklm} & \sim  n(n^2+5) \d^{(5)}{}_{(ijklm)}-2(3n^2-20) \d_{(ijk}\delta_{lm)}  \,, \cdots
\end{align} \noindent
up to numerical coefficients dependent on $n$. The $d$-tensors vanish when their order 
is larger than $n$. The $d$-tensors are totally symmetric and are orthogonal to all other $d$-tensors of different order.
For instance, for the fourth-order tensor this means
\be
d{}_{ijkl} \delta_{ij}=0 \,, \quad
d{}_{ijkl} d{}_{ijk}=0  \,. 
\ee \noindent
Thus, the maximal contraction of the indices of two $d$-tensors of {\it different} order is zero.
Using trace formulas for $\d$-tensors
\be 
	\d{}_{(ijkl)} \d_{ijm}=\tfrac{2}{3}  
\tfrac{(n^2-8)}{n} \d_{klm} \, ,
\ee \noindent
we can find the contraction of two indices for the third-order and fourth-order $d$-tensor
\be 
	d{}_{ijkl} d{}_{ijm}  \sim d{}_{klm} \,.
\ee
Combinations of $d$-tensors provide a basis for the vector space of symmetric invariant polynomials of $SU(n)$.

For $N=3$, the tensor $d_{abc}$ takes the following values
\be
\arraycolsep=6pt\def\arraystretch{1.6}
	\begin{array}{lllll}
	d_{118}=\frac{1}{\sqrt{3}} &
	d_{228}=\frac{1}{\sqrt{3}}  &
	d_{338}=\frac{1}{\sqrt{3}} &
	d_{888}=\frac{-1}{\sqrt{3}}  &
	\\
	d_{448}=\frac{-1}{2\sqrt{3}}  &
	d_{558}=\frac{-1}{2\sqrt{3}} &
	d_{668}=\frac{-1}{2\sqrt{3}}  &
	d_{778}=\frac{-1}{2\sqrt{3}}  &
	\\
	d_{146}=\frac{1}{2}  &
	d_{157}=\frac{1}{2}  &
	d_{247}=-\frac{1}{2} &
	d_{256}=\frac{1}{2}  &
	\\
	d_{344}=\frac{1}{2}  &
	d_{355}=\frac{1}{2}  &
	d_{366}=-\frac{1}{2}  &
	d_{377}=-\frac{1}{2} &
	\end{array}
\ee
The anti-symmetric tensor $f_{abc}$ takes the following values
\be
\arraycolsep=6pt\def\arraystretch{1.6}
	\begin{array}{lll}
	f_{123} = 1 &
	f_{147} = \frac{1}{2} &
	f_{156} = -\frac{1}{2} \\
	f_{246} = \frac{1}{2} &
	f_{257} = \frac{1}{2} &
	f_{345} = \frac{1}{2}\\
	f_{367} = -\frac{1}{2} &
	f_{458} = \frac{\sqrt{3}}{2}&
	f_{678} = \frac{\sqrt{3}}{2}
	\end{array}
\ee

\section{Primary fields}
Here, we write down the full primary operators corresponding to the quasi-primary operators in \sct{sec:su2} and \sct{sec:su3}. Note that all fields are defined only up to an overall normalization factor.
\subsection*{$N=2$ primaries}
The primary field of \equ{n2p61} is
\begin{align}
P_{6,1}= \big(\no{\partial^3 J^a \partial J^a} &-\no{ \partial^2 J^a \partial^2 J^a} - \tfrac{1}{10}\no{\partial^4 J^a  J^a}\big)+ \tfrac{15288}{16465}  \no{\partial^2 T T}  + 
 \tfrac{5838}{16465} \no{\partial T \partial T}+ \nn \\
& + 
\tfrac{ 653}{16465} \partial^4 T- \tfrac{56}{135} \partial^2 Q_4  + 
 \tfrac{112}{45}  \no{Q_4 T} -  \tfrac{22176}{16465}  \no{T T T}\,.
 \end{align}
\subsection*{$N=3$ primaries}
The primary completion of the quasi-primary field $Q_4$ is
\be
\label{An3p4}
	P_4 =  Q_4 - \tfrac{ 9}{31} \partial^2 T+ \tfrac{30}{31}  \no{T T}\,.
\ee
The spin $5$ primary field is
\be
\label{An3p5}
	P_5 = Q_5 + \tfrac{6}{17} \partial^2 Q_3 - \tfrac{42}{85} \no{Q_3 T}  \,.
\ee 
The bilinear spin $6$ primary is
\be
\label{An3p61}
	P_{6,1} =  Q_{6,1} - \tfrac{231}{620}  \no{\partial^2 T T} - \tfrac{21}{310}  \no{\partial T \partial T} + \tfrac{21}{20} \no{Q_4 T} + 
	\tfrac{79}{155} \partial^4 T - \tfrac{49}{40} \partial^2 Q_4 + \tfrac{147}{310} \no{T T T} \,.
\ee
The trilinear spin $6$ primary is
\be
\label{An3p62}
	P_{6,2} =  Q_{6,2}   - \tfrac{ 3}{5} \partial^3 P_3 + \tfrac{24}{5} \no{\partial P_3 T}  - 
 \tfrac{36}{5} \no{\partial T P_3} \,.
\ee 

\section{Algebra of symmetric product orbifold CFT}
The most straightforward way to find the spin and multiplicity of the generators of the symmetry algebra for the symmetric product orbifold theory is by looking at its vacuum character. Let us denote the chiral vacuum character of a seed theory by 
\be
\label{seed}
	\chi_1(q) = \sum_{m=0}^\infty a_m q^m\,. 
\ee
Then the vacuum character of the $N$'th symmetric product orbifold can be read off from the following plethystic exponential \cite{elliptic}
\be
	\chi(\nu,q) = \prod_{m=0}^{\infty} \frac{1}{(1- \nu q^m)^{a_m}}= \exp\(\sum_{k=0}^{\infty} \frac{1}{k} \chi_1(q^k)\nu^k \)\,.
\ee
Expanding the exponential in the RHS gives a series in powers of $\nu$, we get:
\begin{align}
\label{symmProdN}
	\chi(\nu,q) &= \sum_{N=0}^{\infty} \chi_N(q) \nu^N \nn \\
	&= 1 + \chi_1(q) \, \nu + \frac{{\chi_1(q)}^2 + \chi_1(q^2)} {2} \, \nu^2 +
\frac{\chi_1(q)^3 + 3 \chi_1(q) \chi_1(q^2) + 2 \chi_1(q^3)}{6} \, \nu^3 + \cdots\,.
\end{align}
We can find the vacuum character for the $N$'th symmetric orbifold CFT by reading off
the coefficients of $\nu^N$. 
In our case, the seed theory is the single boson theory whose chiral character is given by
\be
\label{seedboson}
	\chi_1(q) = \sum_{m=0}^\infty  a_m q^m = \prod_{n=1}^{\infty} \frac{1}{(1-q^n)}\,.
\ee
Using this expression for $\chi_1$, we can compute the the characters and the corresponding symmetry algebra of the symmetric product orbifold CFT using \equ{symmProdN}. These chiral characters agree with standard results for $S_N$ orbifolds \cite{Bantay:1999}. From these characters, we can compute the spectrum of the algebra for small values of $N$:
\begin{align}
N=2:  \qquad& 1,2,4\,.\nn\\
N=3: \qquad&1, 2, 3, 4, 5, 6^2 \,.\nn\\
N=4: \qquad&1,2, 3, 4^2, 5, 6^3, 7^2, 8^3, 9 \,. \nn\\
N=5:  \qquad& 1, 2, 3, 4^2, 5^2, 6^3, 7^3, 8^5,9^4,10^5,11\,.\nn \\
N=6: \qquad &1, 2, 3, 4^2, 5^2, 6^4, 7^3, 8^6, 9^6, 10^8,  11^7, 12^8, 13\,.\nn
\end{align}
As for the coset case, there can be more generators present. Despite the initial exponential growth in the number of operators with spin, at finite $N$, null states start appearing is the spectrum at some finite value of the spin and thus the algebra truncates.
This is reflected in the  asymptotic growth of the vacuum character:
\be
	n^{-\frac{3}{4}}\exp\({\pi \sqrt{\tfrac{2}{3}  n N}}\)\,.
\ee
which exihibits Cardy growth as $n\rightarrow \infty$.

As $N\rightarrow \infty$, the vacuum character is given by
\be
	\lim_{\nu\rightarrow 1}(1-\nu) \chi(\nu,q) =\prod_{m=0}^{\infty} \frac{1}{(1-q^m)^{a_m}}\,,
\ee
which is again the plethystic exponential of \equ{seedboson}. This can be rewritten as
\be
	\prod_{n = 1}^\infty \frac{1}{(1 - q^{n})^{a_1}}\prod_{\substack{m = 2}}^\infty \; 
\prod_{n = m}^\infty \frac{1}{(1 - q^{n})^{a_m-a_{m-1}}}\,.
\ee
The generators of the infinite $N$ algebra are thus enumerated by the generating function
\be
\label{bosonpart}
	\prod_{n=2}^{\infty} \frac{1}{(1-q^n)}\,.
\ee
along with a spin one field.

\begin{figure}[t!] 
	\begin{center}
		\includegraphics[scale=1]{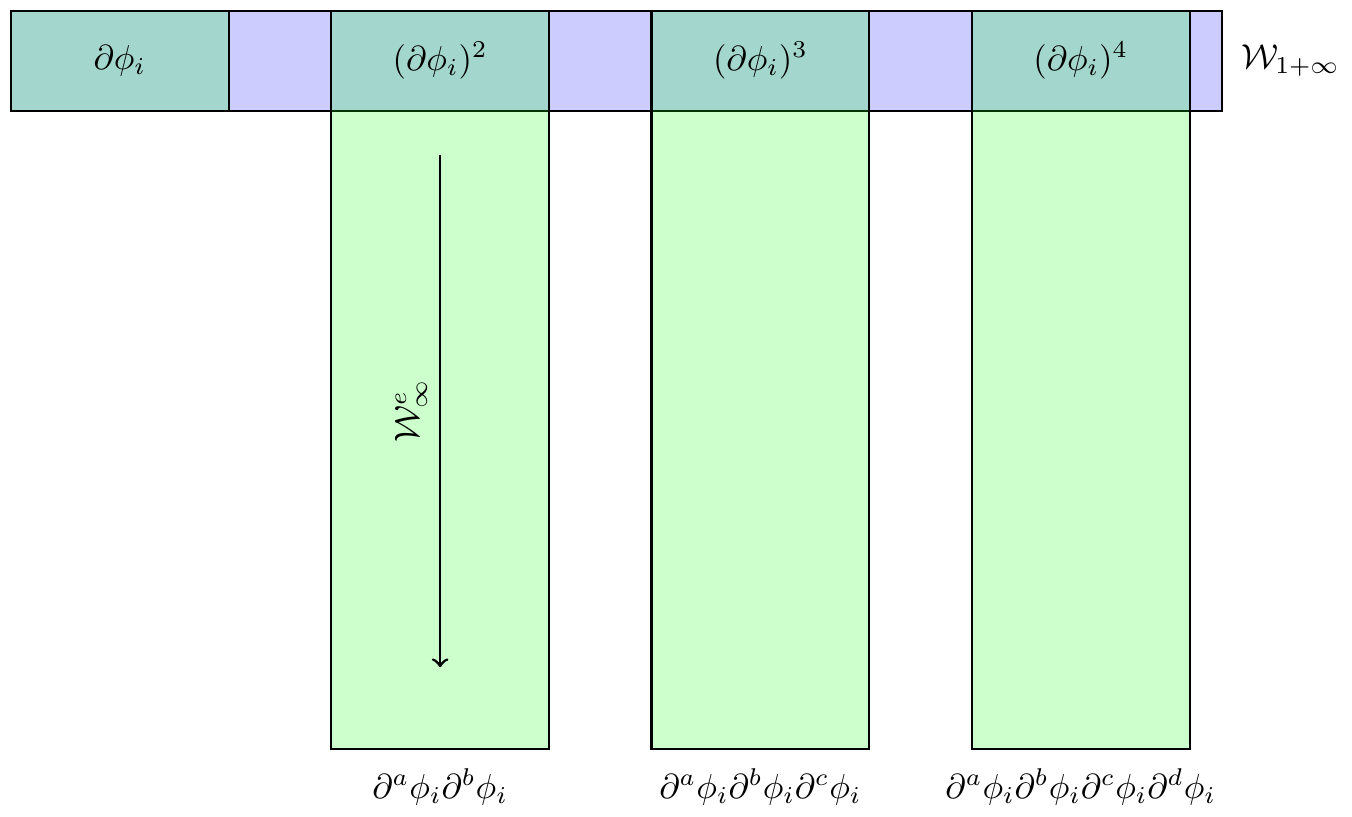}
		\caption{\label{squafig} The higher spin square  is generated by acting by derivatives on the operators in the top row. The second column corresponds to the $\W^e_\infty[1]$ algebra while subsequent columns correspond to its representations.}
	\end{center}
\end{figure}

We now write down the exact form of the currents. In the large $N$ limit, the ``single-particle'' generators for the symmetric product orbifold are symmetrized products of the form
\be\label{spgen}
\sum_{i=1}^{N} \, (\partial^{m_1} \phi_i) \cdots (\partial^{m_p} \phi_i ) \ , \qquad 
m_1,\ldots, m_p\geq 1 \ .
\ee
Because of the symmetrization over $N$, this set of generators is in one-to-one correspondence with the chiral sector of a single boson. Removing the terms that are total derivatives, and in the $N\rightarrow \infty$ limit, they also constitute a set of linearly independent operators.
Out of these generators, the subset of generators of \equ{spgen} of the form
\be\label{bgen}
\sum_{i=1}^{N}\,  (\partial^{m_1} \phi_i)\,  (\partial^{m_2} \phi_i) \ , \qquad m_1,m_2\geq 1 \ , 
\ee
define quasiprimary generators of spin $s=m_1+m_2$, in specific linear combinations and when $s$ is even. In fact, only one independent current  can be constructed at each even spin, meaning that it is not a linear combination of derivatives of lower-spin currents, and there are no independent odd-spin currents. This set of independent currents generate the even spin $\W$-algebra ${\cal W}^{e}_\infty[1]$. 
The generators in \equ{bgen} are of order two, i.e., they are bilinear in the $\phi$s. The currents in \equ{spgen} are of arbitrary order $p\geq 1$.  However, it turns out that the currents of a fixed order $p$, suitably corrected by lower-order terms, form a representation of the wedge algebra of ${\cal W}^{e}_\infty[1]$. This is captured in Fig.~(\ref{squafig}) where currents of a given order correspond to columns. The operators of the symmetric product orbifold  algebra are, therefore, organized into representations of ${\cal W}^{e}_\infty[1]$. The statement that operators of the higher spin square can be organized in representations of ${\cal W}^{e}_\infty[1]$ is captured by the following identity:
\be
\label{VIfull}
\prod_{n=1}^\infty \frac{1}{(1- q^n)} = 1+\sum_{p=1}^{\infty} b^{({\rm wedge}) [\lambda=1]}_{([0^{p-1},1,0,\ldots,0];0)} \,,
\ee
where $b^{({\rm wedge}) [\lambda=1]}_{([0^{p-1},1,0,\ldots,0];0)} $ denotes the wedge character of the $([0^{p-1},1,0,\ldots,0];0)$ representation of ${\cal W}^{e}_\infty[1]$. The $b^{({\rm wedge}) [\lambda=1]}_{([0^{p-1},1,0,\ldots,0];0)}  $ character is defined in \equ{bwedge1}.

The LHS of \equ{VIfull} is the normalized partition function for a single boson. Combinatorially, the LHS is just the generating function for the number of ways one can partition an integer varying from $1$ to $\infty$ into an arbitrary number of parts. Each term in the sum on the RHS is number of ways one can partition an integer into exactly $p$ parts. In terms of the operators in \equ{spgen}, this is the spin $s$ of the operator, varying from $p$ to $\infty$, being partitioned into $m_1,m_2,\ldots,m_p$ at fixed $p$. An alternate way to organize the operators of the higher spin square is in terms of representations of ${\cal W}_{1+\infty}[0]$.


  
\end{document}